\documentstyle[12pt]{article}

\def\section#1{\vspace{2\baselineskip}
		\noindent{\bf #1}\\
		}

\def\subsection#1{\vspace{\baselineskip}
		\noindent{\bf #1}\\
		}

\def\h{\hbox to .5cm{\hfill}}
\def\hof{\hbox to .15cm{\hfill}}
\def\hi{\hbox to .2cm{\hfill}}
\def\htf{\hbox to .35cm{\hfill}}

\def\ha#1{\n\hbox to .6cm{\n {#1}\hfill}}
\def\hb#1{\indent\hbox to .7cm{\n {#1}\hfill}}
\def\hc#1{\indent\hbox to .7cm{\hfill}\hbox to 1.1cm{\n {#1}\hfill}}
\def\hd#1{\indent\hbox to 1.8cm{\hfill}\hbox to 1.4cm{\n {#1}\hfill}}
\def\va{\vskip .2truecm}
\def\n{\noindent}
\def\no{\noindent}
\def\bV{{\bf V}}

\def\balpha{{\bf \alpha}}

\def\bbeta{{\bf \beta}}

\def\theta{\vartheta}

\def\half{{1\over 2}}

\def\third{{1\over 3}}

\def\twothird{{2\over 3}}

\def\fourth{{1\over 4}}

\def\threefourth{{3\over 4}}

\def\fifth{{1\over 5}}

\def\twofifth{{2\over 5}}
\def\threefifth{{3\over 5}}
\def\fourfifth{{4\over 5}}

\def\beq{\begin{equation}}
\def\eeq{\end{equation}}
\def\beqn{\begin{eqnarray}}
\def\eeqn{\end{eqnarray}}

\def\bS{{\bf S}}
\def\mod#1{{\rm (mod~2)}}

 \font\cmss=cmss10 \font\cmsss=cmss10 at 7pt
\def\IZ{\relax\ifmmode\mathchoice
 {\hbox{\cmss Z\kern-.4em Z}}{\hbox{\cmss Z\kern-.4em Z}}
 {\lower.9pt\hbox{\cmsss Z\kern-.4em Z}}
 {\lower1.2pt\hbox{\cmsss Z\kern-.4em Z}}\else{\cmss Z\kern-.4em Z}\fi}

\begin{document}
\rmfamily
\begin{titlepage}
\setcounter{page}{1}
\rightline{OHSTPY-HEP-T-96-013}
\rightline{DOE/ER/01545-680}
\rightline{March 1996}

\begin{center}

\vspace{3\baselineskip}
 {\Large \bf Grand Unified Theories From Superstrings}
\vfill
{\large Gerald B. Cleaver\footnote{E-mail address:
  gcleaver@pacific.mps.ohio-state.edu}\\}
\vspace{.10in}
 {\it Department of Physics, The Ohio State University\\
Columbus, Ohio 43210-1106  USA\\}
\vspace{.20in}
\end{center}
\vfill
\begin{abstract}
I review how traditional grand unified theories, which require adjoint 
(or higher representation) Higgs fields for breaking to the standard model, 
can be contained within string theory. The status (as of January 1996)
of the search for stringy free fermionic three generation SO(10) SUSY--GUT 
models is discussed. Progress in free fermionic classification of both  
SO(10)$_2$ charged and uncharged embeddings and in $N=1$ spacetime 
solutions is presented.
\end{abstract}
\vfill
\begin{center}
{\em Based on talks presented at the Workshop on SUSY Phenomena and SUSY GUTs,
Santa Barbara, California, Dec.~7-11, 1995, and at the Orbis Scientiae, Coral
Gables, Florida, January 25-28, 1996. To appear in the 
Proceedings of Orbis Scientiae, 1996.}
\vfill
\end{center}

\end{titlepage}

\section{SUSY--GUTs and Strings}

Elementary particle physics has achieved phenomenal success in recent
decades, resulting in the Standard Model (SM), 
SU(3)$_{\rm C}\times$SU(2)$_{\rm L}\times$U(1)$_{\rm Y}$, and
verification to high
precision of many SM predictions. However, many aspects of the SM
point to a more fundamental, underlying theory:
\begin{itemize}
\item the SM is very complicated, requiring measurement of some 19 free
      parameters,
\item the SM has a complicated gauge structure,
\item there is a naturalness problem regarding the scale of electroweak 
      breaking,
\item fine-tuning is required for the strong CP problem, and
\item the expected cosmological constant resulting from 
      electroweak breaking is many, many orders of magnitude higher than
      the experimental limit.
%\item the SM provides no unification with gravity.
\end{itemize}  

Since the early 1980's, these issues have motivated investigation of 
Grand Unified Theories (GUTs) 
that would unite SM physics through a single force at higher temperatures. 
Superstring research\cite{schwarz} has attempted to proceed 
one step further and even 
merge SM physics with gravity into a ``Theory of Everything.''

Perhaps the most striking evidence for a symmetry beyond the SM is the
predicted coupling unification not for the SM, but for the minimal 
supersymmetric standard model (MSSM) containing two Higgs 
doublets.\cite{langacker92}
Renormalization group equations applied to
the SM couplings measured around the
$M_{Z^0}$ scale predict MSSM unification at
$M_{\rm unif}\approx 2.5\times 10^{16}$ GeV.
%This prediction imposes an extremely strong constraint on viable string
%theories.
However, this naively poses a problem for string theory,
since the string unification scale has been computed, at tree level, 
to be one order of magnitude higher. That is,
$M_{\rm string}\approx g_s\times 5.5\times 10^{17}$ GeV, 
where the string coupling $g_s\approx 0.7$.\cite{kaplon}
In recent years,
three classes of solutions have been proposed to resolve
the potential inconsistency between $M_{\rm unif}$ and 
$M_{\rm string}$:\cite{dienes1}
\begin{itemize}
\item The unification of the MSSM couplings at
 $2.5\times10^{16}$ GeV should be regarded as a coincidence. 
 $M_{\rm unif}$ could actually be higher as a result of
\begin{enumerate}
     \item SUSY--breaking thresholds,\cite{lightsusy}
     \item non--MSSM states between 1TeV and $M_{\rm unif}$,\cite{faraggig}
     \item non--standard hypercharge normalization 
(a stringy effect),\cite{hypercharge,fermi2} or
     \item non-perturbative effects.\cite{nonpert}
\end{enumerate}
\item $M_{\rm string}$ could be lowered by string threshold 
effects,\cite{kaplon} or
\item $M_{\rm unif}$ and $M_{\rm string}$ remain distinct: 
      there is an effective GUT theory between the two scales.
      MSSM couplings unify around $10^{16}$ GeV and run
      with a common value to the string scale.
\end{itemize}

I have been investigating this third possibility. 
The rationale for this research has been further strengthened recently by
findings suggesting that stringy GUTs and/or 
non--MSSM states between 1TeV and $M_{\rm unif}$ are the only
truly feasible solutions on the list 
(except perhaps for unknown non--perturbative
effects).
Shifts upward in $M_{\rm unif}$ from SUSY--breaking and/or
non--standard hypercharges appear too small to resolve the conflict
and string threshold effects in quasi--realistic models consistently
increase $M_{\rm string}$ rather than lower it.\cite{dienes1}

The birth of string GUTs occurred in 1990, initiated in a  paper by 
D.~Lewellen.\cite{lewellen90} wherein Lewellen constructed
a four--generation SO(10) 
SUSY--GUT built from the free fermionic\cite{kawai87a,antoniadis88} string.
This quickly inspired analysis of constraints on and properties of generic
string GUTs.\cite{font90,ellis90,schwartz90}  Following this, 
string GUT research essentially laid dormant until searches for more 
phenomenologically viable GUTs commenced in 1993 and 1994. 
Initial results during this 
second stage of string GUTs seemed to suggest
that three generation string--derived GUTs were fairly simple to build
and were numerous in number.\cite{cleaver1,fermi1} However, eventually
subtle inconsistencies became evident in all these models. 
The methods used to supposedly yield exactly 
three chiral generations were inconsistent with worldsheet
supersymmetry (SUSY)
and, relatedly, unexpected tachyonic fermions were found in the models.
The desire to produce three generation SUSY-GUTs consistent with
worldsheet SUSY spurred the current stage of string GUT 
research.\cite{cleaver2,fermi2,madrid1,tye95,erler96,faraggi94}

\section{String GUTs and Ka\v c--Moody Algebras}

Besides being the possible answer to the
$M_{\rm unif}/M_{\rm string}$ inconsistency,
string GUTs possess several distinct traits not found in 
non--string--derived GUTs.  
First, string--derived models can explain the origin of the 
extra (local) U(1), R, and discrete symmetries often invoked 
{\em ad hoc.}~in non-string GUTs to significantly restrict superpotential 
terms.\cite{raby1}. The extra symmetries in string models tend to
suppress proton decay and provide for a generic natural mass hierarchy, 
with usually no more than one generation obtaining mass
from cubic terms in the superpotential. All string GUTs have upper limits to
the dimensions of massless gauge group representations that can appear
in a given model. Further, the number of copies of each allowed 
representation is also constrained; there are relationships between
the numbers of varying reps that can appear. 
These features suggest the opportunity for much interplay
between string and GUT model builders.

At the heart of string GUTs are Ka\v c-Moody (KM) algebras, the
infinite dimensional extensions of Lie algebras.\cite{kac83}
(See Table 1.)
A KM algebra may be generated from a Lie algebra 
by the addition of two new elements to the Lie algebra's 
Cartan subalgebra (CSA), $\{ H^i\}$. These new
components are referred to as the ``level'' $K$ and the 
``scaling operator'' $L_0$.
 $K$ forms the center of the algebra, {\em i.e.~}it commutes
with all other members.  Therefore,
$K$ is fixed for a given algebra in
a given string model and is normalized 
to a carry a positive, integer value when the
related Lie algebra is non-abelian.  
$L_0$ appears automatically in a string model as 
the zero-mode of the energy--momentum operator.
These new elements transform the finite
dimensional Lie algebra of CSA and non-zero roots  
$\{H^i, E^{\bf \alpha} \}$ 
into an infinite dimensional algebra,
 $\{K,\, L_0,\, H^i_m,\, E^{\bf\alpha}_m \}$, 
by adding a new indice $m\in\IZ$ to the old elements.
A KM algebra is essentially
an infinite tower of Lie algebras, each distinguished by its $m$-value.

These KM algebras conspire with
conformal and modular invariance
({\em i.e.}~the string self--consistency requirements) 
to produce tight constraints on string GUTs.
There are three generic string--based constraints on gauge groups and 
gauge group reps.
The first specifies the highest allowed level $K_i$ for the
$i^{\rm th}$ KM algebra in a consistent string theory.
The total internal central charge, $c$, 
from matter in the non-supersymmetric sector
of a heterotic string must be 22. The contribution, $c_i$, to this  
from a given KM algebra is a function of the level $K_i$ of the algebra, 
\begin{equation}
c = \sum_{i} c_i = 
    \sum_{i} \frac{K_i {\rm dim}\,{\cal L}_i}{K_i + \tilde{h}_i}\leq 22\,\, .
\label{coninv}
\end{equation}
 dim$\, {\cal L}_i$ and $\tilde{h}_i$ are, respectively, 
the dimension and dual Coxeter of the associated Lie algebra, ${\cal L}_i$.
Eq.~(\ref{coninv}) places upper bounds of 55, 7, and 4, respectively,
on permitted levels of SU(5), SO(10), and E$_6$ KM 
algebras.\cite{font90,ellis90}

Once an acceptable level $K$ for a given KM algebra has been chosen,
the next constraint specifies what Lie algebra reps could potentially
appear. 
Unitarity requires that if a rep, $R$, is to be a primary
field, the dot product between its highest weight, $\lambda^R$, and the
highest root of the KM algebra, $\Psi$, must be less than or equal to K.
\begin{equation}
K\geq \Psi\cdot\lambda^R\,\, .
\label{unitar}
\end{equation}
%where $n_i$ are the Dynkin labels of the highest weight reps of the
%associated Lie algebra $\lal$,and $m_i$ are the related co-marks. 
For example only the $1$, $10$, $16$, and
$\overline{16}$ reps can appear for  SO(10) at level 1. (See table 2.)
For this reason adjoint Higgs require $K\geq 2$ for SO(10) or 
any other KM algebra.

Masslessness of a heterotic string state requires that the total conformal
dimension, $h$, of the non-supersymmetric sector of the state equal one.
Hence the contribution $h_R$ coming from 
rep $R$ of the KM algebra can be no greater than one. For a fixed level $K$, 
$h_R$ is a function of the quadratic Casimir, $C_R$, of the rep,
\begin{equation}
h_R =  \frac{C_r/\Psi^2}{K+\tilde{h}}\,\, .
\label{hmassless}
\end{equation}
Requiring
$h_R\leq 1$ presents a stronger constraint than does unitarity.
For instance, although all SO(10) rep primary fields from the singlet up 
through
the 210 are allowed at level 2, only the singlet up through the 54
can be massless. In particular, the 126 cannot be massless unless
$K\geq 5$.

Free fermionic string models impose one additional constraint.\cite{fermi2}
Increasing the level $K$ decreases the
length-squared, $Q^2_{\rm root}$, of a non-zero root of the KM 
algebra by a factor of $K$. 
In free fermionic strings $Q^2_{\rm root}$ 
at level 1 is normalized to 2 for the long roots. Thus,
\begin{equation}
  K Q^2_{\rm root} = 2\,\, .
\label{KQ2}
\end{equation}
A state containing such a root makes a contribution of 
$\frac{Q^2}{2} =\frac{1}{K}$ to $h$. 
Uncharged free fermionic contributions to $h$
are quantized in units of $\frac{1}{16}$ and $\frac{1}{2}$. Thus,
masslessness of gauge bosons constrain K to be a solution of,
\begin{equation}
  1=  \frac{1}{K} + \frac{m}{16} + \frac{n}{2};\quad 
  m,n\in \{0,\,\IZ^+\}\,\,  ,
\label{mgb}
\end{equation}
which limits $K$ to values in the set $\{1,2,4,8,16\}$.

In combination the constraints (\ref{coninv}) and (\ref{mgb})
permit only levels 1, 2, and 4 for SO(10) and E$_6$,
and, in addition to these, also levels 8 and 16 for SU(5). 
One result is that massless 126's can never appear in free
fermionic SO(10) SUSY--GUTs; 16's must serve in their stead.

\section{SUSY--GUTs From Free Fermionic Models}

In light-cone gauge, a free fermionic heterotic string
model\cite{kawai87a,antoniadis88} contains
64 real worldsheet fermions $\psi^m$, where
$1\leq m \leq 20$ for left--moving (LM) fermions and $21\leq m \leq 64$ 
for right--moving (RM). 
$\psi^1$ and $\psi^2$ are the LM worldsheet
superpartners of the two LM  
scalars embedding the transverse coordinates of
four-dimensional spacetime; the
remaining $\psi^m$ are internal degrees of freedom. 

The transformation property of a real fermion $\psi^{m}$ 
around one of the two non-contractible loops of a torus is expressed by
$\psi^{m} \rightarrow - \exp\{\pi \, i\,\alpha_m\}\psi^m$, and
similarly for the other loop if $\alpha_m$ is replaced by
$\beta_m$. 
The  $\alpha_m$ and $\beta_m$ 
are the $m^{\rm th}$ components of
64--dimensional boundary vectors (BVs) 
${\bf \vec{\alpha}}$ and ${\bf \vec{\beta}}$, respectively,
and have values in the range $(-1,1]$.

If $\psi^m$ cannot be paired with another real fermion or if it is
combined with another to form a Majorana fermion (one LM 
and one RM fermion), its phases are periodic or antiperiodic, 
{\em i.e.}~$\alpha_m,\, \beta_m =  0$ or $1$.
If a real LM (RM) $\psi^m$ is paired with another real LM (RM) $\psi^n$
to form a Weyl fermion $\psi^{m,n}\equiv \psi^n + i\psi^m$, the phases may
be complex 
({\em i.e.}~the BV components 
$\alpha_{m,n}\equiv\alpha_m=\alpha_n$ and 
$\beta_{m,n}\equiv\beta_m=\beta_n$ 
may be rational).

A specific model is
defined by (1) a set of BVs  $\{\bf \vec{\alpha}\}$, 
describing various combinations of  
fermion transformations around the two non-contractible loops on the  
worldsheet torus,
and (2) a set of coefficients, 
$\{C({\bf\vec{\alpha}\atop\bf\vec{\beta}})\}$,
 weighing the contributions, 
$Z({\bf\vec{\alpha}\atop\bf\vec{\beta}})$, to the 
partition function, $Z_{\rm ferm}$, 
from the fermions described by each BV pair $(\vec{\alpha},\vec{\beta})$. 
\begin{equation}
Z_{\rm ferm}= \sum_{{\bf\alpha}\in\{ {\bf\alpha}\}\atop
                         {\bf\beta}\in\{ {\bf\beta}\}}
                  C\left({\bf\vec{\alpha}\atop\bf\vec{\beta}}\right)
                  Z\left({\bf\vec{\alpha}\atop\bf\vec{\beta}}\right) \,\, .
\label{pf}
\end{equation}

The weights $C({{\vec{\balpha}}\atop\vec{\bbeta}})$ can be either 
complex or real ($\pm 1$)
phases when either $\vec{\bf \alpha}$ or $\vec{\bf \beta}$ have rational, 
non-integer components,
but only real phases when $\vec{\bf \alpha}$ and $\vec{\bf \beta}$
are both integer vectors. 

Modular invariance requires that 
$\{ {\bf\vec{\alpha}}\}$ and 
$\{ {\bf\vec{\beta}}\}$ be identical sets and that if two vectors,
$\vec{\bf\alpha}^i$ and $\vec{\bf\alpha}^j$, are 
in $\{ \vec{\bf\alpha}\}$ then
so too is their sum, ${{\bf\vec{\alpha}}^i} +{{\bf\vec{\alpha}}^j}$. 
Thus,
$\{ {\bf\vec{\alpha}}\}$ and $\{ {\bf\vec{\beta}}\}$ 
can be defined by choice of 
some $D'$--dimensional set of basis vectors $\{{\bf \vec{V}}_i\}$,
\beq
\vec{\bf \alpha} = \sum_{i=1}^{D'} a_i {\bf V}_i\,\,\,\,\mod{2}\, ,
\,\,\,\,\,\,
\vec{\bf \beta}  = \sum_{i=1}^{D'} b_i {\bf V}_i\,\,\,\,\mod{2}\, .
\label{expand}
\eeq
Modular invariance also dictates the allowed form of the
phase weights:
\beq
 C\left( {\vec{ \balpha} \atop \vec{\bbeta} } \right)
    = (-1)^{s_{\vec{\balpha}}+ s_{\vec{\bbeta}}}
        \exp \{
             \pi i \, \sum_{i,j} b_i
                      \,( k_{i,j} - \half\, \bV_i\cdot \bV_j  )\,
                                 a_j \}  \, ,
\label{cab}
\eeq
where $s_{\vec{\balpha}}$ ($s_{\vec{\bbeta}}$) 
is the spacetime component of $\vec{\balpha}$ ($\vec{\bbeta}$),
while $k_{i,j}$ is rational and in the range $(-1,1]$. 
There are three mutual constraints on $\bV_i$ and $k_{i,j}$:
\begin{equation}
\begin{array}{rlr}
k_{i,j} + k_{j,i} \,\, = & \half\, \bV_i\cdot \bV_j & \mod{2}\, ,  \\
     N_j k_{i,j}  \,\, = & 0 &                        \mod{2}\, ,  \\
k_{i,i} + k_{i,0} \,\, = & - s_i + \fourth\, {\bV_i}\cdot {\bV_i} & \mod{2}\, .
\end{array}
\label{kconst} 
\end{equation}
$N_j$ is the smallest positive integer such that $N_j\bV_j= 0$ (mod 2).

A complex Weyl fermion $\psi^{n,m}$ in a sector $\vec{\bf\alpha}$ 
carries a U(1) charge $Q_{\balpha}(\psi^{n,m})$ proportional to $\alpha_{m,n}$:
\beq
Q(\psi^{n,m})= \alpha_{n,m}/2 + N(\psi^{n,m}) \,\, .
\label{qdef}
\eeq
$N$ is the fermion number operator and has eigenvalues 0, $\pm1$.
%For (non-)periodic fermions, $Q$ 
%has possible values of $(\{0,\pm 1\})$ $\{\pm\half\}$.
Each  ${\bf\vec{\alpha}}$ yields a set of states
that are excitations of the vacuum by various
modes of the real $\{\psi^{m'}\}$ or complex $\{\psi^{m,n}\}$.
These states, therefore, carry differing charge vectors 
$\bf {Q}_{\balpha}$.
Together, the charges of all states in all sectors form a
lattice upon which the roots and weights of an algebra
can be embedded.
The BVs $\bf\vec{\beta}$
contribute a set of GSO operators that project out 
certain states in each sector: for a state in a sector 
$\vec{\bf\alpha} = a_i\,\bV_i$
to survive, its charge vector ${\bf Q}_{\balpha}$ 
must separately satisfy the relation,
\beq
 \bV_j\cdot {\bf Q}_{\balpha} = \left(\sum_i k_{j,i} a_i\right) + s_j
\,\,\,\, \mod{1}\, ,
\label{gso1b}
\eeq
for each basis vector $\bV_j$.

Consider now level--K SO(10) (henceforth denoted SO(10)$_K$) models. 
As the prior section showed,
the only allowed levels are 1, 2, and 4 corresponding to 
$Q^2_{\rm root}= 1,$ 1/2, and 1/4, respectively. 
Level--2 and level--4 algebras require charge lattices of
dimension greater than the rank of SO(10), {\em i.e.}~more
than five associated U(1) charges are required for each embedding. 
For example, the minimal level--2 embedding requires a six--dimensional
charge lattice, with charge vectors for 
the five SO(10) simple roots given by
$(0,0,0,1,0,0)$, 
$(\half,-\half,-\half,-\half,0,0)$, $(0,0,1,0,0,0)$,
$(0,\half,-\half,0,-\half, \half)$, and
$(0,\half,-\half,0, \half,-\half)$.
The extra degree of freedom on the lattice corresponds to
an additional U(1) algebra.
% (denoted U(1)$_{\rm X}$).

Although the total central charge for SO(10)$_2\times$U(1) is 10,
the charge lattice for SO(10)$_2\times$U(1) only yields a central charge of 6
(since each complex fermion contributes 1 to the central charge).
Additional central charge must come from unpairable real fermions (URFs),
{\em i.e.}~real fermions that cannot form Weyl or Majorana
fermions.\cite{kawai87a} 
URFs assume the role of increasing the
central charge without increasing the  number of local U(1) charges.
It is in this manner that free fermions can 
match the effect of increasing the level of a KM algebra.

Lewellen demonstrated that the smallest possible URF set
is formed from 16 real fermions (containing a central charge of 8).
This set can contribute half of its central charge
to realize the required SO(10)$_2\times$U(1) central charge of 10. 
(Existence of a remaining URF central charge of $4= 8 - 4$ denotes
the presence of a discrete symmetry among the URFs.)
Lewellen's SO(10)$_2$ embedding presents
an example of how free fermionic representations
of higher level KM algebras involve both charged and uncharged sectors.
All of the ``second stage'' attempts at three generation SO(10)$_2$ 
models\cite{cleaver1,fermi1}
involved both the minimal (six--dimensional) SO(10)$_2\times$U(1)
charge embedding and the minimal $c=8$ URF set. 
One direction of my current research is to proceed beyond these minimal
embeddings and comprehensively classify and 
investigate the further possible SO(10)$_2$ charged and uncharged embeddings.
Each physically inequivalent choice of charged and uncharged embeddings should
define a new class of SO(10)$_2$ models. 
When my investigation of
SO(10)$_2$ models is complete, I will proceed on with similar treatment of
SO(10)$_4$. 
One important requirement for all such models is that they have $N=1$ spacetime
SUSY, which is the topic of the next section.

\section{Classes of N=1 Spacetime SUSY Models}

In $D$-dimensional heterotic free fermionic models, 
the $3(10-D)$ real internal LM fermions (henceforth denoted 
by $\chi^I$ rather than $\psi^m$)
non-linearly realize a worldsheet 
SUSY through a  supercurrent of the form\cite{antoniadis87,kawai87a}
\beq
T_{\rm F} = \psi^{\mu}\partial X_{\mu} + f_{IJK}\chi^I\chi^J\chi^K\, .
\label{scharge}
\eeq 
The $f_{IJK}$ are the structure constants of a semi-simple Lie algebra
$\cal L$ of dimension $3(10-D)$.  
Four--dimensional models can involve
any one of the three 18-dimensional Lie algebras: SU(2)$^6$, 
SU(2)$\otimes$SU(4), and SU(3)$\otimes$SO(5). 
When $T_{\rm F}$ is transported around the non-contractible loops 
on the worldsheet, it must transform identically as $\psi^{\mu}$ does: 
periodically for spacetime fermions and antiperiodically for 
spacetime bosons. This requirement severely constrains the BVs in consistent 
models. Since $f_{IJK}\chi^I\chi^J\chi^K$ must transform
as $\psi^{\mu}$ does,
each BV necessarily represents an automorphism (up to a minus sign) of the 
chosen algebra.

The simplest such four--dimensional modular invariant model 
is non--supersymmet-ric.
Its single basis vector is the all--periodic $\bV_0$; therefore it
contains only the sectors $\bV_0$ and $\vec{0}\equiv\bV_0 + \bV_0$
(the all antiperiodic sector).
The graviton, dilaton, antisymmetric tensor, and spin--1 gauge particles
all originate from the $\vec 0$ sector.
Each of the three possible worldsheet SUSY choices for Lie algebra
allows various possibilities for an additional basis vector ${\bf S}_i$
that both satisfies the automorphism constraint and contributes
massless gravitinos. 
Every $\{ \bV,\, \bS_i \}$ set generates an $N=4$ supergravity model.
Additional basis vectors (with related GSO projections) 
must be added to reduce the number of spacetime supersymmetries below four. 
Ref.~\cite{dreiner89b} showed that neither 
SU(2)$\otimes$SU(4) nor SU(3)$\otimes$SO(5) algebras can be used to 
obtain $N=1$ spacetime SUSY.
This work also presented two examples of different basis vector combinations 
(one being the NAHE set of LMs)\cite{faraggix}
 that can yield $N=1$ for SU(2)$^6$, while it
revealed one situation where presence of a specific basis vector 
forbids $N=1$. 

I have finished the work initiated in \cite{dreiner89b}.
That is, I have {\it completely} classified the sets of LM
BVs that can produce exactly $N=1$ spacetime SUSY
(and $N=4$, 2, and 0 spacetime SUSY solutions in the process).  
SU(2)$^6$ is necessarily the supercurrent's Lie algebra,
which gives (\ref{scharge}) the form of,
\beq
T_{\rm F} = \psi^{\mu}\partial X_{\mu} 
             + i\sum_{J=1}^6 \chi^{3J}\chi^{3J+1}\chi^{3J+2}\, .
\label{scharge2}
\eeq 
Each fermion triplet $(\chi^{3J},\, \chi^{3J+1},\, \chi^{3J+2})$ 
represents the three generators of the $J^{th}$ SU(2). 
$N=1$ spacetime is only possible if the generators for each SU(2) are
written in the non--Cartan--Weyl basis of ($J_3$, $J_1$, and $J_2$).

An automorphism of SU(2)$^6$ is
the product of inner automorphisms for the separate SU(2) algebras and
an outer automorphism of the whole SU(2)$^6$ product 
algebra.\cite{antoniadis88,dreiner89b}
The only inner automorphism for an individual SU(2) that could yield
a massless gravitino corresponds to one fermion in a triplet being 
periodic and the other two being antiperiodic.
An outer automorphism can be expressed as an element of the permutation group 
$P_6$ that mixes the SU(2) algebras.\cite{dreiner89b}
The elements of $P_6$ can be resolved into factors of disjoint commuting
cycles. These fit into eleven classes defined by the different
possible lengths, $n_k$, of the cycles in the permutation 
such that $\sum_k n_k = 6$. The set of these eleven classes 
(with a set of lengths written as $n_1\cdot n_2\cdots n_i$) is 
\beqn
{\bf n} &\in \{&1\cdot1\cdot1\cdot1\cdot1\cdot1,\,\,\,\, 
     2\cdot1\cdot1\cdot1\cdot1,\,\,\,\, 2\cdot2\cdot1\cdot1,\,\,\,\,
     2\cdot2\cdot2,
\label{perms}\\
     & &3\cdot1\cdot1\cdot1,\,\,\,\, 3\cdot2\cdot1,\,\,\,\, 
             3\cdot3,\,\,\,\,   4\cdot1\cdot1,\,\,\,\, 4\cdot2,\,\,\,\, 
             5\cdot1,\,\,\,\, 6 \hbox to 0.4cm{\hfill}\}\,.
\nonumber
\eeqn
The first element in this set, $1\cdot1\cdot1\cdot1\cdot1\cdot1$, 
is the $P_6$ identity element, while
$2\cdot1\cdot1\cdot1\cdot1$ is the class with 
cyclic permutation between two SU(2) algebras (which two
is indicated by each class member's $J$ subscripts).
For example,
\beq
2_{1,2}\cdot1\cdot1\cdot1\cdot1:\quad 
(\chi^3,\, \chi^4,\, \chi^5),\, \leftrightarrow  
(\chi^6,\, \chi^7,\, \chi^8)\, . 
\label{perm-b}
\eeq
Similarly, an element of the $2\cdot2\cdot1\cdot1$ class  
permutes two separate pairs of algebras, {\em e.g.}
\beqn
2_{1,2}\cdot2_{3,4}\cdot1\cdot1:\quad(\chi^3,\, \chi^4,\, \chi^5)
         & \leftrightarrow&  (\chi^6,\, \chi^7,\, \chi^8)\, ,\\
(\chi^9,\, \chi^{10},\, \chi^{11}) 
         & \leftrightarrow&  (\chi^{12},\, \chi^{13},\, \chi^{14})\, .
\label{permc}
\eeqn

Of the eleven permutation classes, only those six 
involving an even number of disjoint permutations correspond to
BVs that can yield massless gravitinos.\cite{dreiner89b}
The other five would produce gravitino BVs that cannot satisfy all
requirements of (\ref{kconst}).
The six distinct gravitino BVs are listed in Table 3.
(Note that, as with any BV, a $\IZ_n$ twisted
gravitino generator contains components of the form ${2a\over n}$ where
$a$ and $n$ are relative primes in at least one component.)
I have studied each gravitino generator and applied
all consistent combinations of unique GSO
projections to it.\cite{cleaver2}
I have determined how many of the initial $N=4$ spacetime SUSYs
survive various combinations of GSO projections.
My findings can be summarized as follows:
\begin{enumerate}
\item Only left-moving $\IZ_2$, $\IZ_4$, and $\IZ_8$ twists
that correspond to automorphisms of SU(2)$^6$
are consistent with $N=1$ in free fermionic models.
All other LM $\IZ_n$ twists obviate $N=1$.
Thus, neither gravitino generators ${\bf S}_5$ and 
${\bf S}_7$
(both containing $\IZ_6$ twists), nor ${\bf S}_{10}$ 
(containing $\IZ_{10}$ twists)
can produce $N=1$ spacetime SUSY.
${\bf S}_5$ and ${\bf S}_7$ only result in $N=4$, 2, or 0, whereas
${\bf S}_{10}$ yields $N=4$ or 0.
\item  $N=1$ spacetime SUSY is possible for ${\bf S}_1$, ${\bf S}_3$, 
and ${\bf S}_9$.
Six general categories of GSO projection sets
lead to $N=1$ for ${\bf S}_1$, while 
three do for ${\bf S}_3$, and one does for ${\bf S}_9$.
The GSO projections in all of these sets originate
from LM BVs with the above--mentioned 
$\IZ_2$, $\IZ_4$, and $\IZ_8$ twists.
\end{enumerate}

I have fully 
classified the ways by which the number of spacetime supersymmetries
in heterotic free fermionic strings
may be reduced from $N=4$ to the phenomenologically preferred $N=1$.
This means that the set of LM BVs
in any free fermionic model with claimed $N=1$ spacetime SUSY
must be reproducible from one of the three specific
gravitino sectors in the set 
$\{ {\bf S}_1, {\bf S}_3, {\bf S}_9\}$,
combined with one of my LM BV sets whose GSO projections 
reduce the initial $N=4$ to $N=1$.
The only variations 
from my BVs that true $N=1$ models could have
(besides trivial reordering of BV components)
are some component sign changes that I have shown
do not lead to physically distinct models.

Prior to my present SO(10)$_2$ research,
only the gravitino generator $\bS_1$
had been used in $N=1$ models.
Reduction to $N=1$ spacetime SUSY had always
been accomplished through GSO projections 
from the NAHE set of LM BVs.\cite{faraggix}
Thus, my new $N=1$ solutions should be especially useful for
model building when the NAHE set may be inconsistent with other 
properties specifically desired in a model.
This, indeed, appears to be the situation with regard to
current searches for consistent three generation free fermionic
SO(10) level--2 models, at least when Lewellen's original minimal
charged and uncharged embeddings are chosen.

\section{Concluding Comments}

The result of the 1993--1994 second stage of the search for
string--derived three generation SO(10) SUSY--GUTs was essentially
a free fermionic no--go theorem for a particular choice of 
charged and uncharged SO(10)$_2$ embeddings that was combined with 
the standard gravitino generator, ${\bf S}_1$. 
Free fermionic string GUT research has in 1995 advanced to a more 
mature stage, with classification
of non-minimal charged and uncharged embeddings now underway. 
Further, complete classification of all directions to obtaining 
$N=1$ spacetime SUSY has been completed.
Relatedly, new classes of SO(10)$_2$ free fermionic 
models are now under investigation.
In parallel fashion, SO(10)$_4$ models will also be examined. 
If three generation free fermionic SO(10) SUSY--GUT models do exist,
they will eventually be found through the systematic search
now in operation.

%G.C.~wishes to thank the organizers of {\sc Orbis Scientiae 1996},
%in particular Behram N.~Kursunoglu, for 
%producing such a stimulating and enjoyable conference.  

\section{Note added:}

Since completion of this paper in March, 1996, 
significant advancement has been made in the more general field of
higher--level group theoretic embeddings.\cite{dienes96}
(The free fermionic approach is one method by which such 
embeddings can be realized.)
The recent work discussed in \cite{dienes96}
has led to full classification of group
theoretic embeddings of SU(5), SU(6), SO(10), and E$_6$ in strings.
One result of this is a generalized proof that heterotic
strings cannot yield massless $\bf 126$ representations of SO(10).
\hfill\vfill\eject

\section{Table 1.~Ka\v c--Moody Algebras --vs-- Lie Algebras}

\no\underline{LIE ALGEBRA with rank $l$:}
%\vskip 0.4truecm

\begin{itemize}
\item FINITE dimensional algebra
\end{itemize}

%\vskip 0.5truecm
%\vskip 0.5truecm

\beqn
\left[H^i,H^j\right] &~=~& 0;\,\,\,\, i,\,\, j\in \{ 1,\, 2,\, \dots\, l\}
\nonumber\\
\left[H^i,E^{\bf\alpha} \right] &~=~&\alpha(H^i) E^{\bf\alpha}
\nonumber\\
\left[E^{\bf\alpha},E^{\bf\beta}\right] &~=~& 
 \left\{ \begin{array}{ll}
 \epsilon({\bf\alpha,\beta}) E^{\bf\alpha +\beta,}
         &\mbox{if ${\bf \alpha +\beta}$ is a root;}\\
 {2\over\bf\alpha^2} {\alpha\cdot H}, 
         &\mbox{if ${\bf\alpha+\beta} = 0$;}\\
  0,     &\mbox{otherwise.}
        \end{array}
\right.
\eeqn

\vskip .6truecm
\no\underline{AFFINE KAC-MOODY ALGEBRA with rank $l+2$:}

\vskip 0.4truecm
\begin{itemize}
\item New elements in CSA are ``LEVEL'' $K$ (center of group) and
 ``scaling/energy operator'' $L_0$
\item INFINITE dimensional algebra:\hbox to 1truecm{\hfill}$m,\,\,n\in\IZ$
\end{itemize}
%\vskip 0.4truecm
%\vskip 0.1truecm
%\vskip 0.5truecm

\beqn
\left[H^i_m,H^j_n\right] &~=~& K m\delta^{ij}\delta_{m,-n};\,\,\,\, 
i,\,\, j\in \{ 0,\, 2,\, \dots\, l+1\}
\nonumber\\
\left[H^i_m,E^{\bf\alpha}_n\right] &~=~& \alpha(H^i_0) E^{\bf\alpha}_{m+n}
\nonumber\\
\left[E^{\bf\alpha}_m,E^{\bf\beta}_n\right] &~=~& 
\left\{ \begin{array}{ll}
\epsilon({\bf\alpha,\beta}) E^{\bf\alpha +\beta}_{m+n},
         &\mbox{if ${\bf \alpha +\beta}$ is a root;}\\
{2\over\bf\alpha^2} [{\alpha\cdot H_{m+n}} + Km\delta_{m,-n}] , 
         &\mbox{if ${\bf\alpha+\beta} = 0$;}\\
0,       &\mbox{otherwise.}
        \end{array}
\right.
\nonumber\\
\left[K,H^p_m\right] &~=~& [K,E^{\bf\alpha}_m] = 0 
\nonumber\\
\left[L_0,H^p_m\right] &~=~& -m H^p_m
\nonumber\\ 
\left[L_0,E^{\bf\alpha}_m\right] &~=~& -m E^{\bf\alpha}_m
\nonumber
\eeqn 
\hfill\vfill\eject

%\section{Table 2.~Unitary Massless Gauge Group Reps}

%\input tab2.tex
%  defs below needed for Table 1
\def\bbox#1{\parbox[t]{1.0 in}{{#1}}}
\def\bbreak{\vfill\break}
\def\rep#1{$\bf{#1}$}
\def\ph{$\phantom{h=}$}
\def\pho{$\phantom{0}$}
\def\phn{$\phantom{1}$}

\section{Table 2.~Potentially Massless Unitary Gauge Group Reps}

\begin{tabular}{c|c|c|c|c}
       ~& $k=1$ & $k=2$ & $k=3$ & $k=4$\\
\hline
\hline
SU(5) 
        &
        \bbox{$c=4$\bbreak
         rep\ph {\pho}h\bbreak
         {\pho\rep{5}} \ph2/5\bbreak
         {\rep{10}} \ph 3/5}
         &
        \bbox{$c=48/7$\bbreak
         rep\ph {\pho}{\pho}h\bbreak
           {\pho\rep{5}} \ph 12/35\bbreak
           {\rep{10}} \ph 18/35\bbreak
           {\rep{15}} \ph {\phn}4/5\bbreak
           {\rep{24}} \ph {\phn}5/7\bbreak
           {\rep{40}} \ph 33/35\bbreak
           {\rep{45}} \ph 32/35}
      &
        \bbox{$c=9$\bbreak
         rep\ph {\pho}{\pho}h\bbreak
           {\pho\rep{5}} \ph {\phn}3/10\bbreak
           {\rep{10}} \ph {\phn}9/20\bbreak
           {\rep{15}} \ph {\phn}7/10\bbreak
           {\rep{24}} \ph {\phn}5/8\bbreak
           {\rep{40}} \ph 33/40\bbreak
           {\rep{45}} \ph {\phn}4/5\bbreak
           {\rep{75}} \ph {\phn}1}
      &
        \bbox{$c=32/3$\bbreak
         rep\ph {\pho}{\pho}h\bbreak
           {\pho\rep{5}} \ph {\phn}4/15\bbreak
           {\rep{10}} \ph {\phn}2/5\bbreak
           {\rep{15}} \ph 28/45\bbreak
           {\rep{24}} \ph {\phn}5/9\bbreak
           {\rep{40}} \ph 11/15\bbreak
           {\rep{45}} \ph 32/45\bbreak
           {\rep{50}} \ph 14/15\bbreak
           {\rep{70}} \ph 14/15\bbreak
           {\rep{75}} \ph {\phn}8/9\bbreak
                      \ph}    
        \\
\hline
SO(10) &
        \bbox{$c=5$\bbreak
         rep\ph {\pho}h\bbreak
           {\rep{10}} \ph 1/2\bbreak
           {\rep{16}} \ph 5/8}
         &
        \bbox{$c=9$\bbreak
         rep\ph {\pho}h\bbreak
           {\rep{10}} \ph  9/20\bbreak
           {\rep{16}} \ph 9/16\bbreak
           {\rep{45}} \ph 4/5\bbreak
           {\rep{54}} \ph 1}
          &
        \bbox{$c=135/11$\bbreak
         {\phn}rep\ph {\pho}{\phn}h\bbreak
           {\phn}{\rep{10}} \ph  {\phn}9/22\bbreak
           {\phn}{\rep{16}} \ph 45/88\bbreak
           {\phn}{\rep{45}} \ph {\phn}8/11\bbreak
           {\phn}{\rep{54}} \ph 10/11\bbreak
           {\rep{120}} \ph 21/22\bbreak
           {\rep{144}} \ph 85/88}
         & 
        \bbox{$c=15$\bbreak
         {\pho}rep\ph {\pho}{\phn}h\bbreak
           {\phn\rep{10}} \ph {\phn}3/8\bbreak
           {\phn\rep{16}} \ph 15/32\bbreak
           {\phn\rep{45}} \ph {\phn}2/3\bbreak
           {\phn\rep{54}} \ph {\phn}5/6\bbreak
           {\rep{120}} \ph {\phn}7/8\bbreak
           {\rep{144}} \ph 85/96\bbreak
           {\rep{210}} \ph {\pho}1\bbreak
                       \ph}
            \\
\hline
\end{tabular}

\def\phx{$\phantom{*}$}
\section{Table 3.~Distinct Free Fermionic Gravitino Boundary Vectors}

\no
\underbar{\hbox to 2.4truecm{\bf BV Class\hfill}}
\hskip .2truecm
\underbar{\hbox to 9.0truecm{\bf Gravitino Boundary Vectors\hfill}}
\hskip .2truecm
\underbar{\hbox to 3.0truecm{\bf Allowed SUSY}}

\no
\hbox to 2.4truecm{$1\cdot1\cdot1\cdot1\cdot1\cdot1$\hfill}
\hskip .2truecm
\hbox to 1.1truecm{\hfill ${\bf S}_1$ \hfill $=$}
\hskip .2truecm
\hbox to 9.0truecm{$\{ 1,1\quad (1; 0, 0)^6 \}$ \hfill}\hfill
\hskip .2truecm
\hbox to 3.0truecm{4,{\phx}2,{\phx}1,{\phx}0\hfill}

\va
\no
\hbox to 2.4truecm{$2\cdot2\cdot1\cdot1$\hfill}
\hskip .2truecm
\hbox to 1.1truecm{\hfill ${\bf S}_3$ \hfill $=$}
\hskip .2truecm
\hbox to 9.0truecm{$\{ 1,1\quad (0,1; -\hat{\half}, \hat{\half})^2\quad
                                      (1; 0, 0)^2 \}$ \hfill}\hfill
\hskip .2truecm
\hbox to 3.0truecm{4,{\phx}2,{\phx}1,{\phx}0\hfill}

\va
\no
\hbox to 2.4truecm{$3\cdot1\cdot1\cdot1$\hfill}
\hskip .2truecm
\hbox to 1.1truecm{\hfill ${\bf S}_5$ \hfill $=$}
\hskip .2truecm
\hbox to 9.0truecm{$\{ 1,1\quad
                 (\hat{\third},1; -\hat{\twothird}, 0,0,\hat{\twothird})\quad
                                      (1; 0, 0)^3 \}$ \hfill}\hfill
\hskip .2truecm
\hbox to 3.0truecm{4,{\phx}2,{\phx}\phantom{1,}{\phx}0\hfill}

\va
\no
\hbox to 2.4truecm{$3\cdot3$\hfill}
\hskip .2truecm
\hbox to 1.1truecm{\hfill ${\bf S}_7$ \hfill $=$}
\hskip .2truecm
\hbox to 9.0truecm{$\{ 1,1\quad
                 (\hat{\third},1; -\hat{\twothird}, 0,0,\hat{\twothird})^2
                                                              \}$ \hfill}\hfill
\hskip .2truecm
\hbox to 3.0truecm{4,{\phx}2,{\phx}\phantom{1,}{\phx}0\hfill}

\va
\no
\hbox to 2.4truecm{$4\cdot2$\hfill}
\hskip .2truecm
\hbox to 1.1truecm{\hfill ${\bf S}_9$ \hfill $=$}
\hskip .2truecm
\hbox to 9.0truecm{$\{ 1,1\quad
   (0,\hat\half, 1; -\hat{\threefourth}, -\hat{\fourth},
                     \hat{\fourth},  \hat{\threefourth})\quad
                                      (0,1; -\hat{\half}, \hat{\half})
                                                              \}$ \hfill}\hfill
\hskip .2truecm
\hbox to 3.0truecm{4,{\phx}2,{\phx}1,{\phx}0\hfill}

\va
\no
\hbox to 2.4truecm{$5\cdot1$\hfill}
\hskip .2truecm
\hbox to 1.1truecm{\hfill ${\bf S}_{10}$ \hfill $=$}
\hskip .2truecm
\hbox to 9.0truecm{$\{ 1,1\quad
   (\hat{\fifth},\hat{\threefifth},1; -\hat{\fourfifth},-\hat{\twofifth},0,0,
                                      \hat{\twofifth},  \hat{\fourfifth})\quad
                                      (1; 0, 0)\}$ \hfill}\hfill
\hskip .2truecm
\hbox to 3.0truecm{4,{\phx}{\phx}\phantom{2,1,}{\phx}0\hfill}
\hfill\vfill\eject

\bibliographystyle{unsrt}

\end{document}